\documentclass[submission,copyright,creativecommons]{eptcs}
\providecommand{\event}{ICLP 2022} 

\usepackage{iftex}

\ifpdf
  \usepackage{underscore}         
  \usepackage[T1]{fontenc}        
\else
  \usepackage{breakurl}           
\fi
\usepackage{amsmath}
\usepackage{graphicx}
\usepackage{multirow}
\usepackage{url}

\newcommand{\ruleo}{\;\hbox{:-}\;}

\begin{document}

\title{A Preliminary Data-driven Analysis of Common Errors Encountered by Novice SPARC Programmers}

\author{Zach Hansen
\institute{University of Nebraska Omaha\\Nebraska, USA}
\email{zachhansen@unomaha.edu}
\and
Hanxiang Du 
\institute{University of Florida\\Florida, USA}
\email{h.du@ufl.edu}
\and 
Wanli Xing
\institute{University of Florida\\Florida, USA}
\email{wanli.xing@coe.ufl.edu}
\and
Rory Eckel 
\institute{Texas Tech University\\Texas, USA}
\email{rory.eckel@ttu.edu}
\and
Justin Lugo
\institute{MRC LLC\\USA}
\email{jlug331221@gmail.com}
\and 
Yuanlin Zhang
\institute{Texas Tech University\\Texas, USA}
\email{y.zhang@ttu.edu}
}


\def\titlerunning{Common SPARC Errors}
\def\authorrunning{Hansen et al.}

\maketitle

\begin{abstract}
Answer Set Programming (ASP), a modern development of Logic Programming, enables a natural integration of Computing 
with STEM subjects. This integration addresses a widely acknowledged challenge in K-12 education, and early empirical results on ASP-based integration are promising. Although ASP is considered a simple language when compared with imperative programming languages, programming errors can still be a significant barrier for students. This is particularly true for K-12 students who are novice users of ASP. Categorizing errors and measuring their difficulty has yielded insights into imperative languages like Java. However, little is known about the types and difficulty of errors encountered by K-12 students using ASP. To address this, we collected high school student programs submitted during a 4-session seminar teaching an ASP language known as SPARC. From error messages in this dataset, we identify a collection of error classes, and measure how frequently each class occurs and how difficult it is to resolve. 
\end{abstract}


\section{Introduction}
The emerging need to provide computing education for K-12 (kindergarten to 12th grade) education calls for an integration of computing with STEM (Science, Technology, Engineering and Mathematics) subjects \cite{stemLaw2015,nsf2018}. Computational, scientific, technological, engineering, and mathematical thinking cannot be seen as isolated subjects \cite{swaid2015bringing}.    
The (USA) National Science \& Technology Council Committee on STEM Education (2018) recommends making ``computational thinking an integral element of all education,'' in its five-year (2018 – 2023) strategic plan for STEM education. While substantial efforts have been made to integrate computing with K-12 STEM education in the last decade, the field is still in an early stage of development. 


Answer Set Programming (ASP), a modern development of Logic Programming, has been recently proposed as a basis
for integrating computing and STEM subjects, and the initial results have been promising \cite{YuenRZ19,zhang,ZhangWBMS19}. 
This approach is desirable for a number of reasons. First, ASP focuses on the explicit use of logic in solving problems \cite{GelK14}, and logical reasoning is a common foundation underlying both STEM subjects and Computing. Hence, ASP enables a natural yet deep integration of Computing with STEM subjects. Second, ASP arguably offers essential support for {\em abstraction} and {\em programming}, two of the core computing concepts that are well accepted by the computing education community \cite{k2016k}. Details of these arguments can be found in \cite{YuenRZ19,zhang}. Third, as a purely declarative language, it is considered more natural for students to learn. In fact, in the 1980s, there was extensive study of the use of Logic Programming in promoting the learning of K-12 students (see  \cite{mendelsohn1990programming}). In fact, compared with imperative programming languages, ASP languages usually have simpler constructs.
Furthermore, for STEM problems, ASP provides more 
direct support for representing STEM knowledge and reasoning about it than imperative languages. 
Finally, for logic and reasoning, there is evidence that children ages 11 to 15 already demonstrate substantial knowledge of natural language and the logical use of symbols (see, for example, the work by \cite{piaget1972intellectual} and \cite{vygotsky1987collected}). In fact, this conclusion is supported by the initial ASP-based teaching and learning in middle and high schools \cite{YuenRZ19,zhang}. 

Given the promise of Logic Programming in K-12 teaching and learning, it is worth carrying out more studies on the challenges and issues of this approach. It is well accepted that programming errors can be an obstacle and a significant source of frustration for students \cite{hristova2003identifying}.  
This is an issue for novice users of ASP as well, despite its relative simplicity. 
However, little is known about ASP programming errors, particularly for secondary school students. A better understanding of the errors is expected to improve the teaching and learning of ASP, particularly for the SPARC language studied here. Therefore, the central research question investigated in this paper is: {\bf What is the nature of the programming errors made by high school students using SPARC?} 

Although some of the authors were instructors of several ASP-based modules for secondary schools in the last few years, their experiences are not sufficient to answer this question because the direct interactions between them and the students were limited. Furthermore, neither their memories nor their subjective experiences are sufficiently reliable for further generalization.  
As a result, we decided to answer the question by observing the students' programming ``directly." For this we used the online ASP programming environment onlineSPARC \cite{MarcopoulosZ19}. Specifically, we modified onlineSPARC to record students' editing and testing activities when they wrote ASP programs. We then collected the data from an ASP seminar attended by high school students. 

From this data, we developed a list of error classes that represent a human-centric understanding of the underlying mistakes in the students' programs. We go on to identify the most common of these error classes, and study how difficult each error is to resolve. 
The findings about the most common and difficult errors are expected to directly improve how SPARC is taught in K-12 by enabling educators to address challenging concepts in an informed and systematic manner. Some class time may be spent particularly on frequently occurring or difficult to resolve errors, which will hopefully ease student frustration and accelerate learning. This work will also be useful for constructing better error reporting/correction tools for SPARC programming systems.


In the rest of the paper, we first provide some background in Section~\ref{sec:back} and related work in Section~\ref{sec:related} before we present the experimental design (participants and research setting) in Section~\ref{sec:design}. We then present our analysis of the data in Section~\ref{sec:analysis}. Limitations of our research and the conclusions reached are presented in Section~\ref{sec:limitations} and \ref{sec:conclusion} respectively. 

\section{Background}
We recall here some basics of the ASP language SPARC and programming errors. 
\label{sec:back}
\subsection{SPARC} 
SPARC is an ASP language which allows for the explicit representation of sorts. There are many excellent introduction materials to ASP including \cite{brewka2011answer} and \cite{GelK14}. We will focus on an introduction to SPARC here. The syntax and semantics of SPARC can be found in \cite{BalaiGZ13}, and the SPARC manual and solver are freely available \cite{sparcManual}. 

\begin{figure}[ht]
    \centering
    \fbox{\includegraphics[width=0.7\linewidth]{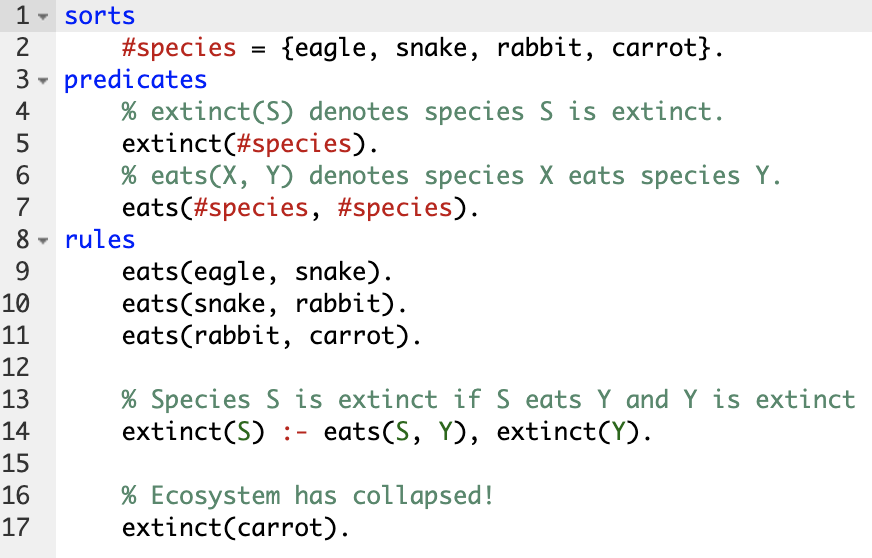}}
    \caption{Food chain problem example in onlineSPARC}
    \label{fig:foodchain}
\end{figure}

To illustrate SPARC, we use the food chain problem, a science topic taught in middle school: consider a chain with carrots, rabbits, snakes and eagles. Typical questions include ``do eagles eat snakes?'' and ``what would happen to eagles if snakes became extinct?'' Fig. \ref{fig:foodchain} shows a SPARC program written inside  onlineSPARC for this problem, which is freely available at \url{http://wave.ttu.edu}. 

A SPARC program consists of three sections: \textit{sorts}, \textit{predicates} and \textit{rules}. The \textit{sorts section} consists of \textit{sort definitions} of the form:
\begin{center} 
\textit{sort\textsubscript{k} = \{object\textsubscript{1}, \ldots,object\textsubscript{n}\},} 
\end{center} 
where \textit{sort\textsubscript{k}} starts with `\#" and $object\textsubscript{1}, \ldots, object\textsubscript{n}$ are called \textit{constants}. For example, line 2 of Fig.~\ref{fig:foodchain} defines valid members of the sort \textit{species}. The \textit{predicates section} consists of \textit{predicate declarations} of the form: 
\begin{center} 
\textit{relationName(sort\textsubscript{1},..,sort\textsubscript{k}),} \end{center} 
where \textit{relationName} is called a \textit{relation} or \textit{predicate} and 
\textit{sort\textsubscript{i}} is the sort of the object involved in the relation.  We say that the predicate \textit{relationName} has $k$ parameters (i.e., a relation over \textit{k} objects). \par
A modeling language like SPARC allows users to formalize knowledge using \textit{atoms} of the form 
\begin{center} 
$p(t\textsubscript{1}, \ldots, t\textsubscript{n})$, 
\end{center} 
where \textit{p} is a relation (predicate) and \textit{t\textsubscript{1}, \ldots, t\textsubscript{n}} are \textit{terms}. A \textit{term} is either a \textit{variable} (an identifier starting with a capital letter) or a member of a sort. The \textit{rules section} consists of knowledge in the form of rules: 
\begin{center} 
$H \ruleo B_1, \dots, B_n$, 
\end{center} 
where each $B_i$ ($0 \leq i \leq n$) is an atom optionally preceded by the negation connective $not$, and $H$ is an atom or empty (in which case the rule is called a {\em constraint}). If $n=0$, the rule is called a {\em fact}.
In Fig. \ref{fig:foodchain}, lines 9-11, 17 are facts. 
Lines beginning with `\%' are comments.
For example, line 13 provides the English description of the knowledge represented by the rule in line 14.

The programming environment onlineSPARC is an online application which allows a programmer to create and edit a SPARC program, to query their program, and to generate the {\em answer sets} of the program. An answer set is a set of 
all the relations of all objects that are believed to be true by a rational agent in terms of the program.
For instance, to answer the question ``Does eagle eat snake?'', we formulate a \textit{query} to the program: \textit{eats(eagle, snake)}? The SPARC solver can derive the answer from the given program: yes. Alternatively, students can click a ``Get Answer Sets'' button, which would display the answer sets of the program. In the case of Fig. \ref{fig:foodchain}, the only answer set includes the facts in lines 9-11, 17 in addition to \textit{extinct(eagle)}, \textit{extinct(snake)} and \textit{extinct(rabbit)}. Alternatively, an error message will be reported if the program contains an error.
\subsection{Programming Errors}
There are four types of programming errors \cite{brain2005debugging}: lexical, syntactic, semantic and conceptual. 
{\em Lexical errors } refer to those related to tokens of a program. A typical lexical error could be a misspelling of a keyword. {\em Syntactic errors} refer to those that inhibit the construction of a proper parse tree. A typical syntactic error could be a missing comma. 
{\em Semantic errors} occur
when syntactically valid language constructs do not have a valid meaning. An example of a semantic error would be the declaration of the same predicate twice. 
In fact, all error messages generated by onlineSPARC when answering a query or finding answer sets belong to one of these three types of errors. 
{\em Conceptual errors} occur when the program, without the errors above, does not behave as intended (produces incorrect answer sets or query responses).  
In this study, we focus exclusively on lexical, syntactic and semantic errors. 
\section{Related Work}
\label{sec:related}
Learning to program is challenging. As such, both educators and researchers want to develop useful strategies to help students. It has been shown that adequate academic preparedness in mathematics and logic helps decrease the difficulty of learning to program \cite{gomes2007learning}. It has also been suggested that programming should be learned in a supported environment where prompt feedback is available \cite{guo2015codechella}. From the programming practice perspective, writing comments, maintaining clear structures, and using meaningful variable names are all considered good habits. Furthermore, debugging ability is critical. Research indicates that students with poor debugging skills 
rarely perform well in their courses \cite{ahmadzadeh2005analysis}. This indicates a need for educators to understand which errors present the greatest challenge to students while debugging. Consequently, considerable efforts have been made to study students’ programming errors. The rationale behind this is quite straightforward. If we know what kinds of errors students tend to encounter and why, educators will be better prepared to develop instructional strategies and useful tools. \par
One source of difficulty for students is understanding error messages and identifying their erroneous code. To address this, Denny et al. (2014) proposed a system to facilitate understanding of Java error messages by combining debugging examples with the appropriate error messages. Closely related work emphasizes producing intelligible error categories from esoteric compiler messages. In this vein, Hristova et al (2003) introduce a collection of error categories identified by surveying computer science educators. This categorization, which spanned syntax, semantic, and logic errors, formed the basis for their interactive tool Expresso. Altadmri and Brown (2015) used this classification system to identify the most common mistakes students encountered while learning Java in a large-scale data set consisting of students from all over the world. Similarly, Qian et al. (2019) studied the common errors encountered by Chinese middle school students learning Python. They found that, in addition to difficulties with Python syntax, comprehension of English versus Chinese punctuation and typing abilities presented significant obstacles to students. \par
In addition to understanding the common errors students encounter, it is also useful to investigate the time spent on fixing these errors. McCall and K\"olling (2019) developed a hierarchical error category and studied the difficulty of different errors. The severity was calculated as the product of error frequency and the time spent on fixing errors, which provided some different ranking results from the frequency alone. Ahmadzadeh et al. (2005) investigated the relationship between students’ performance and the time spent on debugging in an introductory Java course. Their research indicated no correlation between the number of errors and time spent on debugging, yet high-performing students tended to spend less time resolving errors. Denny et al. (2012) also investigated the time spent resolving the most common errors. They found that `Cannot resolve identifier' is the single most common error as well as the longest to resolve. Altadmri and Brown (2015) investigated dynamics of the median time spent on fixing errors for Java learners. They conclude that syntax errors are quick to fix and yet the time spent on fixing these errors is consistent, while semantics and type errors take longer to fix. Additionally, time spent on debugging semantic errors decreases subtly over time, and the time required for type errors fluctuates.

\section{Participants and Research Setting}
\label{sec:design}
To answer the question of ``{\em What is the nature of the programming errors made by high school students using SPARC?}'' we collected data from a seminar for high school students. 

\noindent {\bf Participants.} This study was conducted at a high school in west Texas. A total of 18 students participated in a series of seminars held by one of the authors (Justin Lugo), a graduate student. The students were volunteer participants in a STEM enrichment program, and came from two groups. The first group attended four one-hour lectures, each occurring one week apart. Two example problems were used: the family and food chain problems. A simple example of the food chain problem written by the authors is provided in Fig.~\ref{fig:foodchain}. This problem is an excellent introductory example for integrating STEM and computing learning, as it requires students to understand the dependencies in a food chain; these dependencies translate easily into the language of SPARC. The second group only participated in three lectures due to scheduling conflicts, and did not get a chance to attempt the food chain problem. \par

\medskip
\noindent {\bf Details of the Seminar.} The content was designed by the authors Lugo and Zhang. The first lecture covered the importance of ASP and the family problem. The students were assigned the task of representing their own families using SPARC and learning how to get answer sets from the onlineSPARC environment. This required adapting the sorts and facts of the family problem. During the second lecture, the instructor introduced the concept of predicates and showed how predicates form new relationships such as \textit{mother}. There was no assignment for this lecture, but by the end of the lecture students were expected to define a new relationship within the family problem. This necessitated writing a new predicate. During the third lecture, the students were introduced to rules and variables. The rules \textit{brother} and \textit{sister} were added during the lecture, then the students developed rules for \textit{parent} as homework. The emphasis of this lecture was understanding how to translate rules between English and SPARC. During the fourth lecture, the students were introduced to the food chain problem. The students were given a problem description - they were supposed to model the food chain using provided species and with predicates \textit{eats} and \textit{extinct}. They were expected to apply their SPARC knowledge to a new problem domain. The important concepts introduced over the course of the seminar were modeling problems by identifying objects and their sorts and the relations among these objects, representing them into sort and predicate declarations, formulating English description of the knowledge and writing facts and rules for representing the knowledge, and using variables. Note that due to the limited instruction time, these concepts were addressed at a fairly high level of abstraction. All students practiced their SPARC programming exercises using the onlineSPARC environment. 

\medskip
\noindent {\bf Data Collection.} Whenever a student submits a query or requests all answer sets, onlineSPARC will store the following information: the current program (in the onlineSPARC editor), the timestamp, the student's username, and the result (an error message or list of answer sets). Note that when there is an error in a program, SPARC only reports the first error encountered during program compilation. Hence, for any submission, if the program contains errors, only one error message will be recorded in our dataset. 

\begin{figure}[tbh]
    \centering
     \fbox{\includegraphics[width=0.7\linewidth]{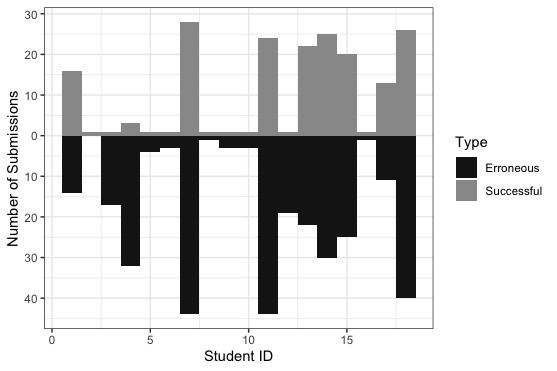}}
    \caption{Number of erroneous and successful submissions per student.}
    \label{fig:REF}
\end{figure}

The final dataset contained 499 total submissions from 18 students. Of these, 313 contained syntax (lexical or syntactic) or semantic errors (62.73\% of the total). This distribution is depicted in Figure \ref{fig:REF}. We refer to these as {\em erroneous} submissions, and all other submissions as {\em successful} even though successful submissions may contain conceptual errors in the answer sets produced. The number of submissions per student ranged from 1 to 72. The mean number of submissions per student was 27.72 with standard deviation 25.23. If we remove the seven students who submitted five or fewer programs, this spread decreases drastically (mean of 43.36 and standard deviation 19.72). The average number of erroneous submissions per student was 17.39 with standard deviation 15.48. 

\section{Data Analysis}
\label{sec:analysis}
In order to understand the nature of the programming errors, we investigated the following: 1) the classes of the errors students made, 2) the frequency of the error classes, and 3) the difficulty of the error classes. 

\subsection{Error Classes}
It is important to draw a distinction between programming errors and error messages.
The two are clearly related, but the literal error messages are often not the error we want to report.
This is particularly true in the case of novice programmers who are not familiar with the syntax and semantics of the language they are using.
To better promote learning, we would like to report the underlying mistake 
instead of its manifestation in the form of error messages.
For example, an error message from onlineSPARC might read: ``predicate gender\_of of arity 2 at line 36, column 5 was not declared.''
We would like to replace such a message with something more general, such as ``Use of an undeclared predicate'' and then provide details about the predicate in question.
This draws attention to the student's mistake, and hopefully diverts them from investigating line 36, where this undeclared predicate was used. 
Even though the {\em error} occurs on line 36, the true issue occurs earlier in the program, where the student omits the necessary predicate declaration.
These observations necessitated the development of a set of general, yet informative error classes.


We base our development of error classes on the data, the SPARC language's syntax and semantic errors, and our judgement. Our judgement focuses on making the classes general and yet potentially related to some common causes of the specific error messages. 
Specifically, we develop our error classes using {\em thematic analysis} \cite{virginia06}.  We first divide the dataset into groups based upon the identification of a common theme (in this case, the cause of each error). Our starting point for developing these classifications was the SPARC manual \cite{sparcManual}. Several of the error messages described in this manual have a singular cause - all instances of \textit{undeclared predicate}, for instance, produced an error message of the form ``predicate \textless token\textgreater  of arity \textless n\textgreater was not declared.'' Most semantic error classes were derived from this manual. 
The remaining error classes were developed iteratively and cooperatively. Three authors examined the error messages in the dataset that did not fit into the SPARC manual categories, and discussed additional categories by which to group the data. 
Each of the two authors label the data using the produced categories earlier and create classes as needed. After the labeling, the three authors will meet and discuss the classes and attempted to form consensus on principles to classify the errors. 
This process was repeated until all error messages in the dataset are labeled by the same class by the two authors.  
A complete list of the resulting classes from the analysis is provided below. It should be noted that this list does not encompass every possible error class for the SPARC language, and instead it represents only the errors present in this dataset.

\medskip
\noindent Syntax errors (including lexical and syntactic):
\begin{enumerate}
    \item\textbf{Sort element name error.} While defining a sort, the programmer uses unacceptable tokens for at least one of its members. 
    Example: In \textit{\#people = \{charlie, bill, Tom\}},
    `Tom' is not a valid member name because it starts with capital `T' and thus is a variable. Note, this error takes precedence over variable error and sort format error (defined later). 
    \item\textbf{Variable error} This error occurs when a programmer incorrectly uses a variable name instead of a constant, or vice-versa. This includes accidental capitalization of constant names outside of sort definitions. For example, the usage of variables instead of sort names in the predicate declaration \textit{eats(X,Y).} Similarly, failure to capitalize a variable causes it to be interpreted as a constant.
    \item\textbf{Missing conjunction.} While defining a rule, the programmer fails to add a conjunctive comma between two atoms. For example: the missing comma between the ending two atoms in the rule \textit{extinct(S):-eats(S, Y) extinct(Y).}
    \item\textbf{Imbalanced structure.} 
    This error occurs when a programmer uses an incorrect or imbalanced syntactical structure. For example: the closing parenthesis in the sort definition \textit{\#gender = \{male, female).}
    \item\textbf{Invalid identifier.} This error occurs when a programmer uses a proper identifier in a syntactically improper place. For example, the misplaced space in the rule 
    \textit{father(buddy, peggy sue).} SPARC does not allow spaces in identifiers. 
    \item\textbf{Incomplete declaration/statement.} This error usually occurs when a programmer writes a declaration or statement without the proper ending period. The error also includes use of an incorrect ending character such as a comma. 
    \item\textbf{obviously incomplete declaration or statement.} This is a subjectively defined subset of the Incomplete declaration/ statement error, where the student is most likely aware the statement or declaration is incomplete. For example, the lack of a body in the sort definition \textit{\#gender =}.
    \item\textbf{English written as code.} This classification covers the event where a programmer writes a declaration or statement in pseudo-code or English instead of the appropriate SPARC code. It also includes forgetting the `\%' notation when writing a comment.
    \item\textbf{Sort format error.} While defining a sort, the programmer either misses the initial `\#' to indicate the sort name, or the subsequent `=' after the sort name.
    \item\textbf{Predicate declaration format error.} While declaring a predicate, the programmer does not follow the predicate declaration syntax. 
    \item\textbf{Rule format error.} The programmer does not follow the rule syntax. For example, the usage of two `:-' operators in \textit{extinct(S):-eats(S,Y):-extinct(Y).}
    \item\textbf{Atom format error.} The programmer does not follow the atom syntax. For example, atom \textit{(X, male)} of the rule \textit{species(S, Y) :- (X, male)}.
    \item\textbf{Misc.} Any syntax error not classified by the above is considered Miscellaneous. This covers cases where a programmer's intention is completely unknown. For example, an extraneous period on a line by itself.
\end{enumerate}
\medskip
\noindent {Semantic errors}: 
\begin{enumerate}
  \item\textbf{Undefined sort.} This occurs when using an undefined sort in a predicate declaration. For example, a programmer might write \textit{gender\_of(\#person, \#gender)} but fail to define the \textit{gender} sort.
  \item\textbf{Redeclared predicate.} This error arises when there are more than one predicates of the same name declared in the predicates section of a program.
  \item\textbf{Undeclared predicate.} This issue appears when a predicate is used in a rule or fact without first being declared in the predicates section. For example, using the predicate \textit{father} in the fact \textit{father(bob, sara)} without declaring it in the predicates section would trigger an error of this class.
  \item\textbf{Invalid predicate argument sort.} This occurs when an argument of an atom does not follow the expected sort of it. For example, assuming we have sort definition \textit{\#people = \{bob, sahra\}} and predicate declaration \textit{father(\#people, \#people)}, the atom \textit{father(bob, sara)} would cause this error because the argument \textit{sara} is not a member of the sort \textit{\#people}.
\end{enumerate}
\subsection{Common Errors}

%

For any erroneous student submission, it is labeled by a unique error class. The {\em frequency} of an error class is the number of submissions labeled by this class. The {\em relative frequency} of an error class is its frequency divided by the total number of erroneous submissions. The commonness of an error class is measured by its frequency and relative frequency. 

Table~\ref{fig:freq} shows the frequency and relative frequency (columns `Freq.' and `Rel. Freq.') of all errors. This represents the entire dataset, with all erroneous submissions from both problems. 
Semantic errors make up 47.6\% of all errors, the other 52.4\% were syntax errors. The three most frequent error classes comprise the majority of all errors (58.8\%). \textit{Incomplete declaration/statement} is the most common error class, followed closely by \textit{Undeclared predicate} and \textit{Undefined sort}.

\begin{table*}[htb]
    \centering
    \begin{tabular}{| p{0.35\linewidth} | p{0.08\linewidth} | p{0.09\linewidth} | p{0.09\linewidth} | p{0.1\linewidth} | p{0.1\linewidth} |}
    \hline
    Error Class & Freq. & Rel. Freq. & Res. Freq. & Time & Attempts \\
    \hline
    Incomplete declaration/statement & 70 & 22.4\% &92.3\% & 182.3 & 3.5\\
    \hline
    Undeclared predicate & 62 & 19.8\% & 73.7\% & 299.6 & 3.6 \\
    \hline
    Undefined sort & 52 & 16.6\% & 81.8\% & 194.2 & 4.1\\
    \hline
    Invalid predicate argument sort & 22 & 7.0\% & 72.7\% & 139.6 & 1.4\\
    \hline
    Sort element name error & 16 & 5.1\% & 50\% & 11 & 1\\
    \hline
    Predicate declaration format error & 15 & 4.8\% & 80\% & 48.6 & 2.3\\
    \hline
    Redeclared predicate & 13 & 4.2\% & 50\% & 180 & 9\\
    \hline
    Atom format error & 12 & 3.8\% & 50\% & 508 & 9\\
    \hline
    Rule format error & 10 & 3.2\% & 75\% & 284.3 & 3.7\\
    \hline
    Malformed structure & 9 & 2.9\% & 75\% & 206.3 & 9.3\\
    \hline
    Missing conjunction & 7 & 2.2\% & 100\% & 64.3 & 2.3\\
    \hline
    English written as code & 6 & 1.9\% & 100\% & 113.7 & 2.3\\
    \hline
    Sort format error & 6 & 1.9\% & 100\% & 254 & 6\\
    \hline
    Invalid identifier & 5 & 1.6\% & 100\% & 54.3 & 1.7\\
    \hline
    Obviously incomplete declaration or statement & 5 & 1.6\% & 100\% & 198.5 & 1.3\\
    \hline
    Misc. & 2 & 0.6\% & 100\% & 26 & 2\\
    \hline
    Variable error & 1 & 0.3\% & 100\% & 30 & 1\\
    \hline
    \end{tabular}
    \caption{Frequency and resolution difficulty of error classes}
    \label{fig:freq}
\end{table*}

Some students made the same mistake many times, while also submitting far more programs than their peers. The raw frequencies depicted in the `Freq.' column of Table~\ref{fig:freq} may, in such cases, create the impression that the given mistake was a common one for all students. The spread of the data necessitates validation of these impressions. We also provide an alternative way to measure how widespread an error is by examining how common an error is for each student. The \textit{most common error count} of an error class is the number of students for whom the error class is their most frequent mistake. The data is shown in Table ~\ref{fig:MCE}. In fact, the frequency and most common error counts coincide in the most common errors: \textit{incomplete declaration/statement}, \textit{undeclared predicate}, and \textit{undefined sort}. This suggests a stability of the most common errors identified.

\begin{table} [ht]
\centering
\begin{tabular}{|c|c|}
\hline
Error Class & Count\\
\hline
Incomplete declaration/statement & 5\\
\hline
Undefined sort & 3\\
\hline
Undeclared predicate & 3\\
\hline
Invalid predicate argument sort & 3\\
\hline
Sort format error & 1\\
\hline
Sort element name error & 1\\
\hline
Missing conjunction & 1\\
\hline
Invalid identifier & 1\\
\hline
\end{tabular}
\caption{The `Count' column represents the number of students for whom `Error Class' was their most frequently made error.}
\label{fig:MCE}
\end{table}
\subsection{Resolution Difficulty}
To determine how difficult each error was for students to resolve, we investigated the frequency of resolution of each error class (Table \ref{fig:freq}, `Freq. Res.' column). Additionally, we provide the average elapsed time and average number of attempts to arrive at a resolution for each error class (Table \ref{fig:freq}, columns Time and Attempts, respectively). A \textit{problem submission sequence} of a student is the sequence of all submissions by the student for this problem. An \textit{occurrence} of an error class in a problem submission sequence is a student submission belonging to this error.  An error occurrence is \textit{resolved} if there exists a later submission where this error is resolved. A student is said to {\em resolve an error} if the program lines causing the error are fixed. If the student introduces a new error that the compiler catches earlier, the error message may change without a resolution being achieved. This represents a new error occurrence for a different error class. As such, it is possible for a single submission to count as an attempt to resolve multiple ongoing error occurrences. An occurrence of an error is called \textit{intermediate} if there exists an earlier error that has the same error message (ignoring information such as line or column number) and is not resolved. An occurrence of an error is \textit{new} if it is not intermediate. \par
The number of submissions (also called \textit{attempts}), not including the first one with the error, required to resolve a new error occurrence can be employed to measure the difficulty of the error. When students did not resolve the error occurrence, we count the number of submissions until the last one in that problem submission sequence as attempts. If the student abandons the problematic lines (e.g., deletes or comments them out), a resolution is not achieved until they return to the lines and fix them. The exception to this is when the problematic lines are unnecessary for achieving the goal of the program - in such a case removing the lines is the right thing to do and counts as a resolution. For each error class, the \textit{frequency of resolution} is the ratio of the number of new occurrences of the error that are resolved to the total number of new occurrences of the error class. Essentially, it is the rate at which students successfully resolve, through a sequence of attempts, the given error. This was our primary measure for how difficult an error was to resolve. \par
A secondary measurement of error class difficulty is the time needed to resolve an error. The \textit{resolution time} of a new error occurrence is approximated by the elapsed time (in seconds) from the new error occurrence to the submission showing the error is resolved. The average resolution time (Table \ref{fig:freq}, column Time) for each error class is defined as the summation of the resolution time for each new occurrence of this error that is ultimately resolved divided by the total number of new resolved occurrences of this error. Finally, we also provide the average number of submissions (Table \ref{fig:freq}, column Attempts) required to achieve resolution. This metric was used in conjunction with elapsed time as a secondary measure of difficulty. Note that occurrences that never get resolved are not included in these metrics. Furthermore, there were four new occurrences where the student took a week-long break from their program; here the length of the break was subtracted from the resolution time. \par
These metrics provide a basis for comparing the difficulty presented by each error class. In conjunction with the relative frequency of errors, they allow us to determine which errors present the largest challenges to students. First, it is evident that semantic errors prove more challenging to resolve than syntax errors. The average resolution frequency (resolved new occurrences divided by total new occurrences) was 72.42\% for semantic errors, compared to 88.14\% for syntax errors. The average number of attempts required to resolve semantic errors (3.375) closely matched that of syntax errors (3.385). However, the average time for resolution of semantic errors was 226.25 seconds, substantially higher than the 160.83 seconds required, on average, for syntax errors. With a lower rate of resolution and a longer time required to achieve it, semantic errors appear to be more challenging. \par
The difficulty of the three most common error classes is of particular interest, since they are so frequently encountered by students. It is interesting to note that the most common error class, \textit{incomplete declaration/statement} has a higher than average rate of resolution (92.3\%). This error can occur for a number of reasons, mainly uncertainty about ending punctuation or accidental omissions, and rarely presents an insurmountable challenge. This indicates that it does not represent widespread conceptual misunderstandings. \par
Conversely, \textit{undeclared predicate} has the lowest resolution frequency of the top three most common error classes (although it is close to average for semantic resolution frequency). The average time for resolution is much higher than the semantic and syntax average, at 299.6 seconds. As such, it presents more of a challenge to students than \textit{Incomplete declaration/statement}, or \textit{undefined sort}.  

\section{Limitations}
\label{sec:limitations}
A primary limitation of this work is the size of the study. The overall body of work is fairly small (499 programs), as is the length of the course (4 seminars) and the number of students (18). 
The advantage of a smaller dataset is that we were able to manually inspect each program. This allowed for careful refinement of the error classes, and accurate identification of the elapsed time and number of attempts for new occurrence resolution. 
This is not to say that manual inspection is a foolproof technique. 

Another limitation is the subjective nature of our qualitative analysis. The error classifications presented in this work were created to help humans understand the underlying cause of each error. This sometimes lacks the precision of an error message produced by a compiler. A good example of this is the class of \textit{obviously incomplete declaration or statement}. While the authors tried to make educated guesses about when a student recognized their program was incomplete as opposed to accidental omission, there was always a degree of subjectivity. Similarly, the decisions about when a resolution was achieved and how long it took sometimes made assumptions about the student's programming behavior. Removing the week-long breaks from the elapsed time is likely safe, but it is unclear if this should apply to smaller gaps as well.
Furthermore, if a student is dealing with multiple simultaneous error occurrences, there is not a definitive way to determine if they are really trying to resolve all the errors or just the one creating the error message. Examples of both are common in our programming experience. \par

Despite the size of the dataset, it still suggests some meaningful patterns in the nature of errors. For example, in Table~\ref{fig:freq}, sort and predicate related errors are significantly more common than other errors. 
This could likely be explained by the fact that much more time was spent on teaching rules in the seminar than on the sort related constructs (i.e., sort and predicate declarations).
Of course, we cannot exclude the possibility that
sort and predicate declarations might indeed cause more errors than rules. 

\section{Conclusions and Discussion} 
\label{sec:conclusion}
This study examines novices' use of the SPARC language. From the data collected from high school students, we identify a list of error classes that reflect the underlying cause of the error from a programmer's perspective. The majority of student errors are attributable to the three most common error classes: \textit{incomplete declaration/statement}, \textit{undeclared predicate}, and \textit{undefined sort}. 
We also study the resolution difficulty for the error classes by using the frequency of resolution, the number of attempts and the elapsed time required for a resolution. According to all three criteria, {\em atom format error} and {\em imbalanced structure} are clearly challenging. The next level of difficulty is {\em rule format error}, {\em redeclared predicate}, {\em undeclared predicate} and {\em undeclared sort}.   

The data-driven analysis offers a set of error classes, their frequency, and their difficulty, which are richer insights than those provided by our own teaching experiences. One example is that imbalanced structures and atom format errors are indeed not as frequent as we thought. But it did not occur to us that they are hard to correct once made until we examined the data. We believe that these sorts of conclusions can provide immediate strategies to address programming errors. For example, one can raise students' awareness of most of these errors and their corresponding error message templates (in terms of the frequency) and design practices (e.g., asking students to correct a bug) in terms of the difficulty of the errors. Sufficient attention should also be paid to the sort and predicate declarations during lecture time. 

A possible avenue of future work could focus on using the error classes established in this project for intelligent error reporting that can suggest corrections. Such software would likely draw inspiration from the Expresso tool~\cite{hristova2003identifying}. Developing software to provide more helpful error messages has the potential to help students move on more quickly when they encounter errors. 
Another, more ambitious, direction would be to extend the methods we have presented here to categorizing logical/conceptual errors. For instance, our teaching experiences suggest that students have difficulty understanding aggregates, specifically that the conditions define a {\em set} of tuples as opposed to a list. However, a data-driven approach would provide more reliable insights into conceptual error frequency and difficulty than our isolated experiences.
Finally, conducting a similar study with a larger body of student work and (possibly) an automated process of determining error class and resolutions could enhance the generalizability of our findings. 

\section{Acknowledgements}
This work is partially supported by NSF grant DRL-1901704. We thank Michael Gelfond and Rocky Upchurch for discussions and help during this project. We gratefully acknowledge the valuable feedback from our reviewers as well.

\bibliographystyle{eptcs}
\bibliography{bib}
\end{document}